\documentclass[aps,prl,revtex4-1,twocolumn,superscriptaddress,showpacs,letter,longbibliography]{revtex4-1}

\usepackage{amssymb}
\usepackage{amsmath}
\usepackage{graphicx}
\usepackage{dcolumn}
\usepackage{bm,textcomp}
\usepackage{subfigure}
\usepackage{braket}
\usepackage{hyperref}

\begin{document}
\title{Freely configurable quantum simulator based on a two-dimensional array of individually trapped ions}
\author{Manuel Mielenz}
\author{Henning Kalis}
\author{Matthias Wittemer}
\author{Frederick Hakelberg}
\affiliation{
Albert-Ludwigs-Universit\"at Freiburg, Physikalisches Institut, Hermann-Herder-Strasse 3, 79104 Freiburg, Germany
}
\author{Roman Schmied}
\affiliation{
University of Basel, Department of Physics, Klingelbergstrasse 82, 4056 Basel, Switzerland
}
\author{Matthew Blain}
\author{Peter Maunz}
\affiliation{
Sandia National Laboratories, P.O. Box 5800 Albuquerque, NM 87185-1082, USA
}
\author{Dietrich Leibfried}
\affiliation{
Time and Frequency Division, National Institute of Standards and Technology, 325 Broadway, Boulder, CO 80305,
USA
}
\author{Ulrich Warring}
\affiliation{
Albert-Ludwigs-Universit\"at Freiburg, Physikalisches Institut, Hermann-Herder-Strasse 3, 79104 Freiburg, Germany
}
\email{ulrich.warring@physik.uni-freiburg.de}
\author{Tobias Schaetz}
\affiliation{
Albert-Ludwigs-Universit\"at Freiburg, Physikalisches Institut, Hermann-Herder-Strasse 3, 79104 Freiburg, Germany
}
\affiliation{Albert-Ludwigs-Universit\"at Freiburg, Freiburg Institute for Advanced Studies, Albertstr. 19, 79104 Freiburg, Germany}
\date{\today}
\begin{abstract}
A custom-built and precisely controlled quantum system may offer access to a fundamental understanding of another, less accessible system of interest. A universal quantum computer is currently out of reach, but an analog quantum simulator that makes the relevant observables, interactions, and states of a quantum model accessible could permit experimental insight into complex quantum dynamics that are intractable on conventional computers. Several platforms have been suggested and proof-of-principle experiments have been conducted. Here we characterise two-dimensional arrays of three ions trapped by radio-frequency fields in individually controlled harmonic wells forming equilateral triangles with side lengths 40 $\mu$m and 80 $\mu$m. In our approach, which is scalable to arbitrary two dimensional lattices, we demonstrate individual control of the electronic and motional degrees of freedom, preparation of a fiducial initial state with ion motion close to the ground state, as well as tuning of crucial couplings between ions within experimental sequences. Our work paves the way towards an analog quantum simulator of two-dimensional systems designed at will.
\end{abstract}
\maketitle

%
\section*{Introduction}
Richard Feynman was one of the first to recognise that quantum
systems of sufficient complexity cannot be simulated on a
conventional computer~\cite{Feynman1982}. He proposed to use a
quantum mechanical system instead. A universal quantum computer (QC)
would be suitable, but practical implementations are
a decade away at best. However, universality is not required to
simulate specific quantum models. It is possible to a custom-build analog quantum simulator (AQS) 
that allows for preparation of fiducial input states, faithful implementation of
the model-specific dynamics, and for access
to the crucial observables.
Simulations on such AQSs could impact a vast variety of research
fields~\cite{Georgescu2014}, i.e., physics~\cite{Schaetz2013},
chemistry~\cite{Lanyon2010}, and
biology~\cite{Fleming2011}, when studying dynamics that is out of
reach for numerical simulation on conventional computers.

Many experimental platforms have been suggested to implement AQSs~\cite{Aspuru-Guzik2012,Houck2012,Bloch2012,Blatt2012}.
Different experimental systems provide certain advantages in
addressing different physics. Results that are not conventionally
tractable may be validated by comparing
results of different AQSs simulating the same problem~\cite{Cirac2012, leibfried_could_2010}.
Over the last two decades, many promising proof-of-principle
demonstrations have been made using photons~\cite{Aspuru-Guzik2012},
superconductors~\cite{Houck2012}, atoms~\cite{Bloch2012}, and trapped
atomic ions~\cite{Blatt2012}. Trapped ions in particular have seen steady
progress from demonstrations with one or two ions~\cite{leibfried_trapped-ion_2002,Friedenauer2008,Schmitz2009,schmitz_arch_2009,Zaehringer2010,Gerritsma2010,matjeschk_experimental_2012}
to addressing aspects of quantum magnets~\cite{Porras2004} with
linear strings of two to 16 ions~\cite{Friedenauer2008,Islam2013} and self-ordered two-dimensional
crystals containing more than 100 ions~\cite{Britton2012}. 
Ions are well suited to further propel the research since they provide
long range interaction and individual, fast controllability with
high precision~\cite{blatt_entangled_2008}. 

Two-dimensional trap-arrays may offer advantages over trapping in a common potential, because they are naturally suited to implement tunable couplings in more than one spatial dimension. Such couplings are, in most cases, at the heart of problems that are currently intractable by conventional numerics~\cite{Cirac2012,Hauke2012}. Our approach is based on surface-electrode structures~\cite{Seidelin2006} originally developed for moving ion qubits through miniaturised and interconnected, linear traps as proposed in~\cite{Wineland1998, Kielpinski2002}. This approach is pursued successfully as a scalable architecture for QC, see, for example~\cite{amini_toward_2010}. For AQSs it is beneficial to have the trapped ion ensembles coupled all-to-all so they evolve as a whole.  This is enabled by our array architecture with full control over each ion. Individual control allows us to maintain all advantages of single trapped ions while scaling the array in size and dimension~\cite{schaetz_towards_2007,Schmied2009,Schmied2011}.

Optimised surface electrode geometries can be found for any periodic
wallpaper group as well as quasi-periodic arrangements, as, for
example Penrose-tilings~\cite{Schmied2009}.
A first step, trapping of ions in two-dimensional arrays of
surface traps, has been demonstrated~\cite{Sterling2014}.
Boosting the strength of interaction to a level comparable to 
current decoherence rates requires inter-ion distances $d$
of a few tens of micrometers. Such distances have been demonstrated in
complementary work, where two ions have been trapped in individually
controlled sites of a linear surface-electrode trap at $d$ between
$30\,\mu$m and $40\,\mu$m. The exchange of a single quantum of
motion, as well as entangling spin-spin interactions have been
demonstrated in this system~\cite{Brown2011,Wilson2014}.
The increase in coupling strength was achieved with a reduction of the ion-surface separation to order $d$ and the concomitant increase in motional heating due to electrical noise.
Recently, methods for reducing this heating by more than two orders
of magnitude with either surface
treatments~\cite{Allcock2011,Hite2012,daniilidis_surface_2014} or
cold electrode
surfaces~\cite{deslauriers_scaling_2006,labaziewicz_suppression_2008,labaziewicz_temperature_2008} have been devised.

Here we demonstrate the precise tuning of all relevant parameters of
a two-dimensional array of three ions trapped in individually
controlled harmonic wells on the vertices of equilateral triangles with side lengths 80 and 40\,$\mu$m. In the latter, Coulomb coupling rates~\cite{Brown2011} approach current rates of decoherence.
Dynamic control permits to reconfigure Coulomb and laser
couplings at will within single experiments. We initialise 
fiducial quantum states by optical pumping, Doppler and
resolved sideband cooling to near the motional ground state.
Our results demonstrate important prerequisites for experimental quantum
simulations of engineered two-dimensional systems.

\section*{Results}

\begin{figure*}[ht!]
\centering
\includegraphics[width=\textwidth]{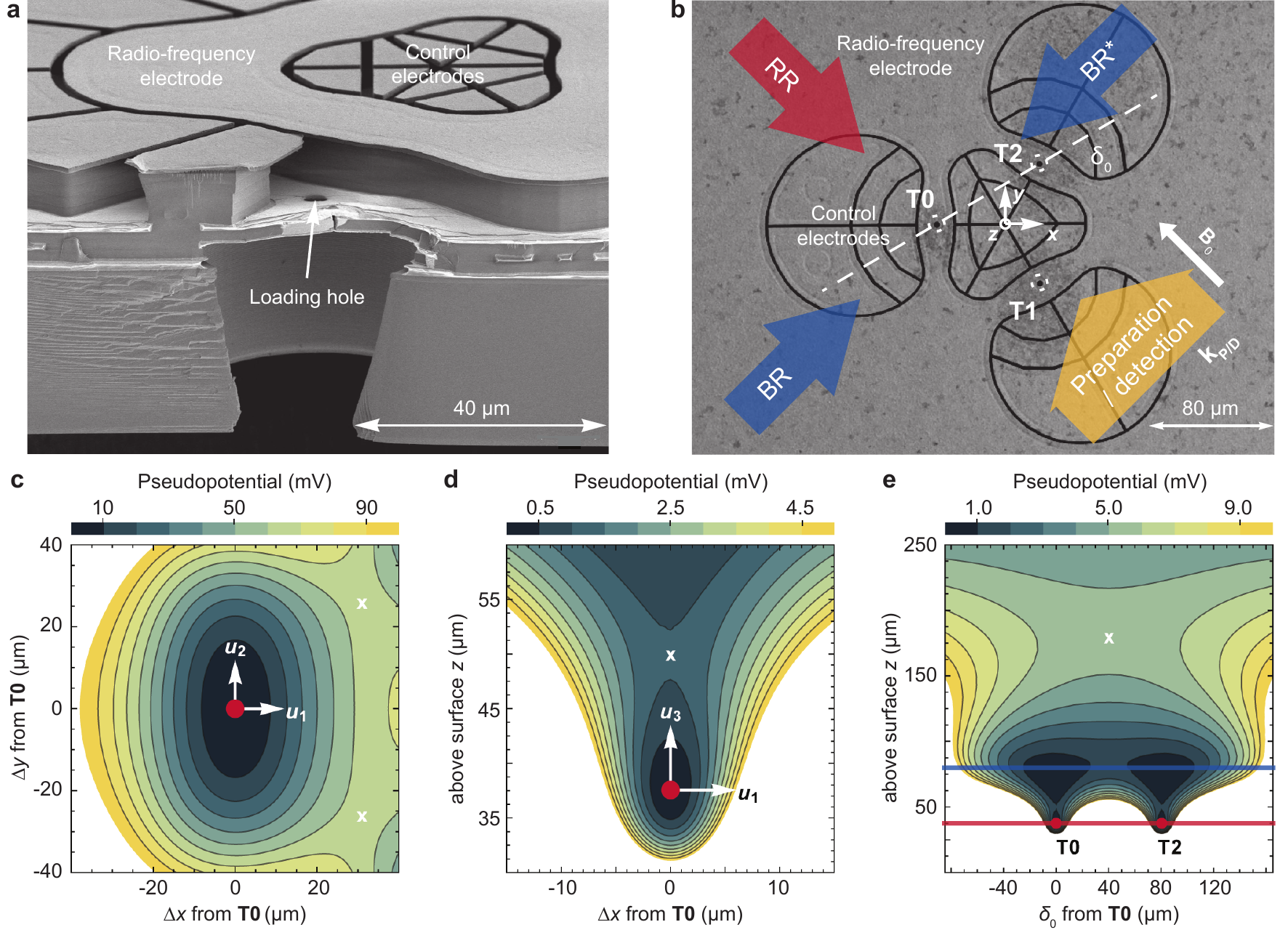}
\caption{\label{fig:TrapSetup} %
  \textbf{Surface-electrode ion trap featuring three individual traps.} %
  (\textbf{a})~Scanning electron microscope (SEM) image of a cleaved copy of our chip. It provides a cross-sectional view vertically through the trap chip (bottom half of image) and a top view of the horizontal, planar trap electrode surface (top half of image) of the 40 $\mu$m array. Buried electrode interconnects as well as the overhangs of electrodes that shield trapped ions from insulating surfaces are exposed in this view.
A loading channel, vertically traversing the chip, collimates a
neutral atom beam from an oven on the backside of the chip. %
  (\textbf{b})~SEM top-view of the 80 $\mu$m array, dark lines indicate gaps between individual electrodes and dashed circles highlight the three trap sites at $\mathbf{T0}$, $\mathbf{T1}$, and $\mathbf{T2}$ that lie 40\,$\mu$m above the electrode plane; corresponding loading holes appear as dark spots.
  A vertical plane connecting $\mathbf{T0}$ and $\mathbf{T2}$ is shown as a dotted line and labelled with $\delta_0$.
  The RF electrode extends beyond the image area and encloses 30 control electrodes grouped into four islands, enabling the control of individual trap sites.
  Laser beams (coloured arrows) are parallel to the chip surface and wave vectors $\mathbf{k}_\textrm{P/D}$ of preparation and detection beams are parallel to the magnetic quantisation field $\mathbf{B}_0$ (white arrow).
  (\textbf{c,~d,~e}) show the pseudopotential $\phi_\textrm{ps}(x, y, z)$ of the 80 $\mu$m array in different planes, trap
sites marked by red dots, motional mode vectors $\mathbf{\hat u}_{j}$ with $j = \{1,2,3\}$ at $\mathbf{T0}$ are
represented by white arrows, and saddle points are illustrated by white crosses.
In (\textbf{e}) the heights of $\mathbf{T0}$, $\mathbf{T2}$ and the ancillary trap are indicated by red and blue lines, respectively.
}
\end{figure*}

\textbf{Trapping and control potentials.}
Our surface ion trap chip is fabricated in similar fashion to that described in~\cite{Tabakov2015} and consists of two equilateral triangular trap arrays with side length of $\simeq 40\,\mu$m and $\simeq 80\,\mu$m, respectively (Fig.~\ref{fig:TrapSetup}a and \ref{fig:TrapSetup}b), both with a distance of $\simeq 40\,\mu$m between the ions and the nearest electrode surface.
The shapes of radio-frequency (RF) electrodes of the arrays are optimised by a linear-programming algorithm that yields electrode shapes with low fragmentation, and requires only a single RF-voltage source for operation~\cite{Schmied2009, Schmied2011}.
%
The two arrays are spaced by $\simeq 5$\,mm on the chip, and only one of them is operated at a given time.
While we achieve similar results in both arrays, the following discussion is focussed on the 80 $\mu$m array.

Three dimensional confinement of $^{25}$Mg$^{+}$ ions is provided by a potential from a single RF electrode driven at $\Omega_\textrm{RF}/(2 \pi) = 48.3$\,MHz with an approximate peak voltage $U_{\textrm{RF}} = 20$\,V.
Setting the origin of the coordinate system at the center of the array and in the surface plane of the chip, the RF potential features three distinct trap sites at $\mathbf{T0} \simeq (-46,\, 0,\, 37)\,\mu$m, $\mathbf{T1} \simeq (23,\,-23\,\sqrt{3},\, 37)\,\mu$m, and $\mathbf{T2} \simeq (23,\,23\,\sqrt{3}, 37)\,\mu$m.
Due to the electrode symmetry under rotations of $\pm 2 \pi/3$ around the $z$-axis (bold symbols with hat denote unit vectors), it is often sufficient to consider $\mathbf{T0}$ only, since all our findings apply to $\mathbf{T1}$ and $\mathbf{T2}$ after an appropriate rotation.
Further, the RF potential exhibits another trap site at $\simeq (0,\, 0,\,81)\,\mu$m (above the center of the array); this `ancillary' trap is used for loading as well as for re-capturing ions that escaped from the other trap sites. %

We approximate the RF confinement at position $\mathbf{r}$ by a pseudopotential $\phi_{\textrm{ps}}(\mathbf{r}) = Q/(4\,m\,\Omega_{\textrm{RF}}^2)\,E_{\textrm{RF}}^2(\mathbf{r})$~\cite{Ghosh1995},
where $Q$ denotes the charge and $m$ the mass of the ion, and $E_{\textrm{RF}}(\mathbf{r})$ is the field amplitude produced by the electrode. Calculations of trapping potentials are based on~\cite{schmied_electrostatics_2010}, utilising the software package~\footnote{R. Schmied, Surface-Pattern software package (2015)}. Equipotential lines of $\phi_{\textrm{ps}}$ are shown in Fig.~\ref{fig:TrapSetup}c--\ref{fig:TrapSetup}e.

Near $\mathbf{T0}$ we can approximate $\phi_{\textrm{ps}}$ up to second order and diagonalise the local curvature matrix to find normal modes of motion described by their mode vectors $\mathbf{u}_1$, $\mathbf{u}_2$, and $\mathbf{u}_3$, which coincide (for the pure pseudopotential) with  $x$, $y$, and $z$; we use $\mathbf{u}_j$ with $j = \{1,2,3\}$ throughout our manuscript to describe the mode vectors of a single ion near $\mathbf{T0}$.
We find corresponding potential curvatures of $\kappa_{\rm{ps, 1}} = 3.4 \times 10^7$\,V m$^{-2}$, $\kappa_{\rm{ps, 2}} = 1.5 \times 10^7$\,V m$^{-2}$, and $\kappa{\rm{ps, 3}} = 1.9 \times 10^7$\,V m$^{-2}$, while mode frequencies can be inferred from these curvatures as $\omega_j = \sqrt{(Q/m)\, \kappa_{\textrm{ps},j}}$, with $j = \{1,2,3\}$: $\omega_{\rm{1}}/(2\pi) \simeq 5.4\,$MHz, $\omega_{\rm{2}}/(2\pi)  \simeq 2.4$\,MHz, and $\omega_{\rm{3}}/(2\pi)  \simeq 3.0$\,MHz. %

To gain individual control of the trapping potential at each site, it is required to independently tune local potentials near $\mathbf{T0}$, $\mathbf{T1}$, and $\mathbf{T2}$ (Methods), i.e., to make use of designed local electric fields and curvatures.
To achieve this, we apply sets of control voltages to 30 designated control electrodes (see Fig.~\ref{fig:TrapSetup}). In the following, a control voltage set is described by a unit vector $\mathbf{\hat{v}_{\textrm{c}}} \equiv (\hat{v}_{\textrm{c},1}, \dots, \hat{v}_{\textrm{c},30})$, with corresponding dimensionless entries $\hat{v}_{\textrm{c},n}$ with $n=\{1,\dots,30\}$, and result in a dimensionless control potential
\begin{equation} 
\hat{\phi}_{\rm{c}} = \sum_{n=1}^{30} \hat{v}_{\textrm{c},n} \, \hat{\phi}_{n}(\mathbf{r})\textrm{,}
\end{equation}
where $\hat{\phi}_{n}(\mathbf{r})$ is the potential resulting when applying $1$\,V to the $n$-th electrode~\cite{Blakestad2011}.
We scale $\hat{\phi}_{\textrm{c}}$ by varying a control voltage $U_{\rm{c}}$ and yielding a combined trapping potential
\begin{equation} 
\phi(\mathbf{r}) = \phi_{\textrm{ps}}(\mathbf{r}) + U_{\rm{c}}\,\hat\phi_{\textrm{c}}(\mathbf{r})\rm{.}
\end{equation}
Bias voltages applied to the control electrodes are, in turn, fully described by $\mathbf{U}_{\rm{c}} = U_{\rm{c}}\,\mathbf{\hat{v}_{\textrm{c}}}$.

To design a specific $\hat\phi_{\rm{c}}$, we consider the second order Taylor expansion for a point $\mathbf{r_0}$ and small displacements $\mathbf{\Delta r}$:
\begin{eqnarray} 
\hat{\phi}_\textrm{c} \left(\mathbf{r}_0 + \mathbf{\Delta r}\right) \simeq
\hat\phi_{\textrm{c}} (\mathbf{r}_0)
+\left[ \partial_k \right]^{\mathbf{T}} \hat{\phi}_{\textrm{c}}(\mathbf{r})|
_{\mathbf{r} = \mathbf{r}_0}\cdot \mathbf{\Delta}\mathbf{r} \nonumber \\
+ \frac{1}{2} \mathbf{\Delta}\mathbf{r}^{\mathbf{T}}\cdot [\partial_k \partial_l] \hat{\phi}_{\textrm{c}}(\mathbf{r})|_{\mathbf{r} = \mathbf{r}_0} \cdot\mathbf{\Delta r}\textrm{,}
\end{eqnarray}
where $\left[ \partial_k \right]^{\mathbf{T}} \hat{\phi}_{\textrm{c}}(\mathbf{r})|
_{\mathbf{r} = \mathbf{r}_0}$ is the local gradient and $[\partial_k \partial_l] \hat{\phi}_{\textrm{c}}(\mathbf{r})|_{\mathbf{r} = \mathbf{r}_0}$ is the traceless and symmetric matrix with indices $k$ and $l$ = $\{x, y, z\}$ that describes the local curvature; square brackets denote vectors/matrices, $\partial$ partial derivatives, and the superscript $\mathbf{T}$ the transpose of a vector. %
We constrain local gradients in their three degrees of freedom (DoF) and local curvatures in their five DoF at $\mathbf{T0}$, $\mathbf{T1}$, and $\mathbf{T2}$, and solve the corresponding system of 24 linear equations to yield~$\mathbf{\hat v}_{\textrm{c}}$. In principle, it would be sufficient to use 24 control electrodes, however, we consider all electrodes and use the extra degrees of freedom to minimise the modulus of the voltages we need to apply for a given effect.

In particular, we distinguish two categories of control potentials, denoted by $\hat\varepsilon$ and $\hat\kappa$, respectively: %
The first category is designed to provide finite gradients and zero curvatures at $\mathbf{T0}$, with zero gradients and curvatures at $\mathbf{T1}$ and $\mathbf{T2}$; for example, $\hat\phi_{\textrm{c}} =
\hat\varepsilon_\textrm{x}$ provides a gradient along $\mathbf{\hat{x}}$ at $\mathbf{T0}$.
Control potentials of the second category are designed to provide zero gradients and only curvatures at $\mathbf{T0}$, while we require related gradients and curvatures to be zero at $\mathbf{T1}$ and $\mathbf{T2}$.
For example, we design $\hat\phi_{\textrm{c}} = \hat\kappa_{\rm{tune}}$, with the following non-zero constrains $\partial_y \partial_y \hat\kappa_{\rm{tune}}(\mathbf{r})|_{\mathbf{r} = \mathbf{T0}} = $ $-\partial_z \partial_z \hat\kappa_{\rm{tune}}(\mathbf{r})|_{\mathbf{r} = \mathbf{T0}} = 0.937\times10^{7}\,$m$^{-2}$ with corresponding $U_{\rm{c}} = U_{\rm{tune}}$.
Linear combination of multiple control potentials enable us, e.g., to locally compensate stray potentials up to second order, to independently control mode frequencies and orientations at each trap site, and, when implementing time-dependent control potentials, to apply directed and phase controlled mode-frequency modulations or mode excitations. %

\textbf{Laser beam setup.} We employ eight laser beams at wavelengths near 280\,nm, from three distinct laser sources~\cite{friedenauer_high_2006}, with wave vectors parallel to the $xy$ plane  (Fig.~\ref{fig:TrapSetup}b) for preparation, manipulation, and detection of electronic and motional states of $^{25}$Mg$^+$ ions.
Five distinct $\sigma^+$-polarised beams (two for Doppler cooling, two for optical pumping and one for state detection) are superimposed, with wave vector $\mathbf{k}_{\textrm{P/D}}$ (preparation/detection) aligned with a static homogeneous magnetic quantisation field $\mathbf{B_0} \simeq 4.65$\,mT (Fig.~\ref{fig:TrapSetup}b). The beam waists (half width at $1/\textrm{e}^2$ intensity) are $\simeq 150\,\mu$m in the $xy$ plane and $\simeq 30\,\mu$m in  $z$ direction, to ensure reasonably even illumination of all three trap sites, while avoiding excessive clipping of the beams on the trap chip.
The two Doppler-cooling beams are detuned by $\Delta \simeq -\Gamma/2$ and $-10\,\Gamma$ with respect to $\ket{\downarrow}$ $\equiv$ $\ket{S_{1/2}, F = 3, m_F = +3}$ $\leftrightarrow$ $\ket{P_{3/2}, F = 4, m_F = +4}$ [with $\Gamma/(2\pi) \simeq 42$\,MHz].
The state detection beam is resonant with this cycling transition and discriminates $\ket{\downarrow}$ from $\ket{\uparrow} \equiv \ket{S_{1/2}, F = 2, m_F = +2 }$, the pseudo-spin states $\ket{\downarrow}$ and $\ket{\uparrow}$ are separated by $\omega_0/(2\pi) \simeq 1681.5$\,MHz. %
The resulting fluorescence light is collected with high numerical aperture lens onto either
a photomultiplier tube (PMT) or an electron-multiplying charge-coupled device (EMCCD) camera. %
We prepare (and repump to) $\ket{\downarrow}$ by two optical-pumping beams that couple $\ket{\uparrow}$ and $\ket{S_{1/2}, F = 3, m_F = +2}$ to states in $\ket{P_{1/2}}$ from where the electron decays back into the ground state manifold and population is accumulated in $\ket{\downarrow}$.
We can couple $\ket{\downarrow}$~to~$\ket{\uparrow}$ via two-photon stimulated-Raman transitions~
\cite{Monroe1995,Wineland1998,Leibfried2003a}, while we can switch between two different beam configurations labelled \mbox{BR*\,+\,RR} with $\mathbf{\Delta k}_{\textrm{x}} \parallel \mathbf{\hat{x}}$ and \mbox{BR\,+\,RR} with $\mathbf{\Delta k}_{\textrm{y}} \parallel \mathbf{\hat{y}}$. The beam waists are $\simeq 30\,\mu$m in the $xy$ plane and $\simeq 30\,\mu$m in $z$ direction.

\textbf{Experimental protocol} We load ions by isotope-selective photoionisation from one of three atomic beams collimated by $4\,\mu$m loading holes located beneath each trap site (Fig.~\ref{fig:TrapSetup}).
We can also transfer ions from one site to any neighbouring site via the ancillary trap by applying suitable potentials to control electrodes and a metallic mesh (with high optical transmission) located $\simeq 7$\,mm above the surface.
Typically, experiments start with 2\,ms of Doppler cooling, optionally followed by resolved sideband cooling, and $\ket{\downarrow}$ preparation via optical pumping.
We use 30 channels of a 36-channel arbitrary waveform generator (AWG) with 50\,MHz update rate~\cite{Bowler2013} to provide static (persistent over many experiments) and dynamic (variable within single experiments) control potentials.
Each experiment is completed by a pulse for pseudo-spin detection of duration $\simeq 150\,\mu$s that yields $\simeq12$ counts on average for an ion in $\ket{\downarrow}$ and $\simeq 0.8$ counts for an ion in $\ket{\uparrow}$.
Specific experimental sequences are repeated 100\,to\,250 times.

\textbf{Operation and characterisation of the trap array.} %
Initially, we calibrate three (static) control potentials $\hat\varepsilon_{\textrm{x}}$, $\hat\varepsilon_{\textrm{y}}$, and $\hat\varepsilon_{\textrm{z}}$ to compensate local stray fields~\cite{Berkeland1998} with a single ion near $\mathbf{T0}$, while we observe negligible effects on the local potentials near $\mathbf{T1}$ and $\mathbf{T2}$ (Methods). Rotated versions of these control potentials are used to compensate local stray fields near $\mathbf{T1}$ and $\mathbf{T2}$. Near each site, we achieve residual stray field amplitudes $\leq 3$\,V\,m$^{-1}$ in the $xy$ plane and $\leq 900$\,V\,m$^{-1}$ along  $z$, currently limited by our methods for detection of micromotion.

With the stray fields approximately compensated, we characterise the trap near $\mathbf{T0}$ with a single ion (Methods). We find mode frequencies of $\omega_1/(2\pi) \simeq 5.3$\,MHz, $\omega_2/(2\pi) \simeq 2.6$\,MHz, and $\omega_3/(2\pi) \simeq 4.1$\,MHz with frequency drifts of about 1.5\,Hz\, s$^{-1}$; mode frequencies and orientations are altered by local stray curvatures on our chip, in particular, $\mathbf{u}_1$ and $\mathbf{u}_3$ are rotated in the $xz$ plane, while $\mathbf{u}_2$ remains predominantly aligned along  $y$. We obtain heating rates for the modes $\mathbf{u}_1$ of 0.9(1)\,quanta\,ms$^{-1}$, $\mathbf{u}_2$ of 2.2(1)\,quanta\,ms$^{-1}$, and $\mathbf{u}_3$ of 4.0(3)\,quanta\,ms$^{-1}$.

\begin{figure}[]
\includegraphics{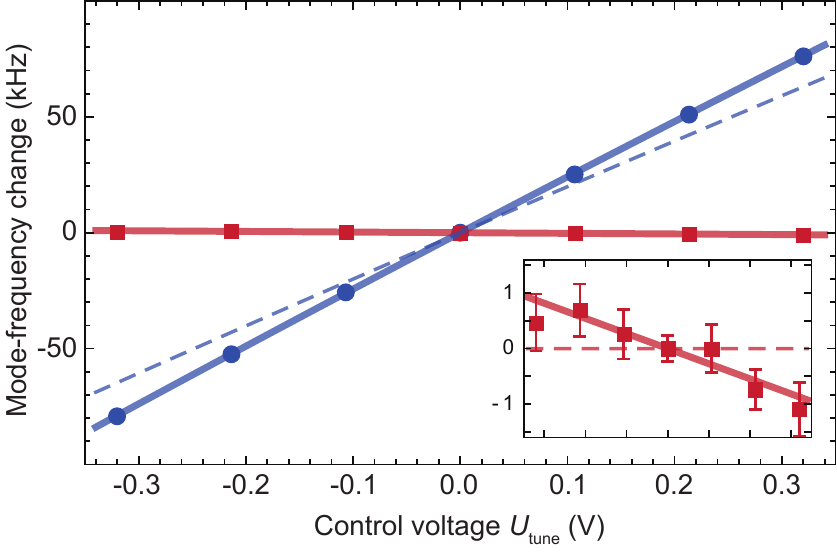}
\caption{\label{fig:ShiftT03}
\textbf{Individual control of mode frequencies.} Frequency change $\Delta\omega_{2}$ probed with a single ion near $\mathbf{T0}$ (blue dots) as a function of $U_{\rm{tune}}$.
The intended control curvature $0.937\times10^{7}$\,m$^{-2}$ (along the $y$ direction) yields the blue/dashed line [cp. equation~(\ref{Eq:DelOme})], while a fit to the data (blue/solid line) returns a control curvature of $1.164(3)\times10^{7}$\,$m^{-2}$.
The residual change of the corresponding mode frequency in kHz of a single ion near $\mathbf{T2}$ (red squares) is shown in the inset.
A fit to these data (red/solid lines) results in a residual control curvature of $-0.012(2)\times10^{7}$\,m$^{-2}$.
Ideally, $\hat\kappa_{\rm{tune}}$ would create no additional curvature at $\mathbf{T2}$ (red/dashed line).
We attribute the difference between designed and measured values to residual ion displacements from $\mathbf{T0}$ and $\mathbf{T2}$.
Each datapoint represents the average of 250 experiments with error bars (for some data smaller than symbols) denoting the s.e.m.
}
\end{figure}

\textbf{Control of mode frequencies and orientations at individual trap sites.}  The ability to control mode frequencies and orientations at each site with minimal effect on local trapping potentials at neighbouring sites is essential for the static and dynamical tuning of inter-ion Coulomb couplings. We experimentally demonstrate individual mode-frequency control using $\hat\kappa_{\rm{tune}}$. To this end we measure local mode frequencies with a single ion near $\mathbf{T0}$ or $\mathbf{T2}$ (Methods). Tuning of about $\pm 80$\,kHz of $\omega_{2}$ near $\mathbf{T0}$ is shown in Fig.~\ref{fig:ShiftT03} as blue data points, accompanied by residual changes of about $\mp 1$\,kHz in the corresponding mode frequency near the neighbouring site $\mathbf{T2}$, depicted by red data points. To infer local control curvatures, we describe the expected detuning $\Delta \omega_{2}$ due to $\hat\kappa_{\textrm{tune}}$ at $\mathbf{T0}$ (analogously at $\mathbf{T2}$) by
\begin{equation}\label{Eq:DelOme}
\Delta \omega_{2} = \sqrt{\omega_{2}^2\,+\,U_{\rm{tune}}\,\left(\frac{Q}{m}\,\partial_y \partial_y \hat\kappa_{\rm{tune}}(\mathbf{r})|_{\mathbf{r}  = \mathbf{T0}} \right)} -
\omega_{2}\textrm{,}
\end{equation}
where we neglect a small misalignment of $\mathbf{u}_2$ from  $y$.
The prediction of equation~(\ref{Eq:DelOme}) is shown as a blue/dashed line in Fig.~\ref{fig:ShiftT03}. The blue/solid line results from a fit with a function of the form of equation~(\ref{Eq:DelOme}) to the data yielding a control curvature of $1.164(3)\times10^{7}$\,m$^{-2}$.
%
The inset magnifies the residual change in frequency near $\mathbf{T2}$. Here, a fit (red/solid line) reveals a curvature of $-0.012(2)\times10^{7}$\,m$^{-2}$. Residual ion displacements of $\Delta z = -2.95(3)\,\mu$m  from $\mathbf{T0}$ and $\Delta z = -2.9(4)\,\mu$m from $\mathbf{T2}$, respectively, suffice to explain deviations between experimentally determined and designed curvature values and are below our current limit of precision locating the ions in that direction.
In future experiments, curvature measurements may be used to further reduce stray fields.

We also implement a dynamic $U_{\rm{tune}}(t)$, to adiabatically tune $\omega_2$ near $\mathbf{T0}$ within single experiments:
We prepare our initial state by Doppler cooling, followed by resolved sideband cooling of mode $\mathbf{u}_2$ to an average occupation number $\bar{n}_2 \simeq 0.1$ and optical pumping to $\ket{\downarrow}$.
In a next step, we apply a first adiabatic ramp from $U_{\rm{tune,A}} = 0$\,V to $U_{\rm{tune,B}}$ between $0$\,V and $2.3$\,V within $t_{\textrm{ramp}}= 7.5\,\mu$s to $120\,\mu$s and, subsequently,  couple $\ket{\downarrow}$ and $\ket{\uparrow}$ to mode $\mathbf{u}_2$ with pulses of \mbox{BR\,+\,RR} tuned to sideband transitions that either add or subtract a quantum of motion. If the ion is in the motional ground state, no quantum can be subtracted and the spin state remains unchanged when applying the motion subtracting sideband pulse. The motion-adding sideband can always be driven, and comparing the spin-flip probability of the two sidebands allows us to determine the average occupation of the dynamically tuned mode~\cite{Leibfried2003a}.
We find that the average occupation numbers are independent of the duration of the ramp and equal to those obtained by remaining in a static potential for $t_{\textrm{ramp}}$, i.e., the motion is unaffected by the dynamic tuning.%

\begin{figure}[h]
  \centering
  \includegraphics{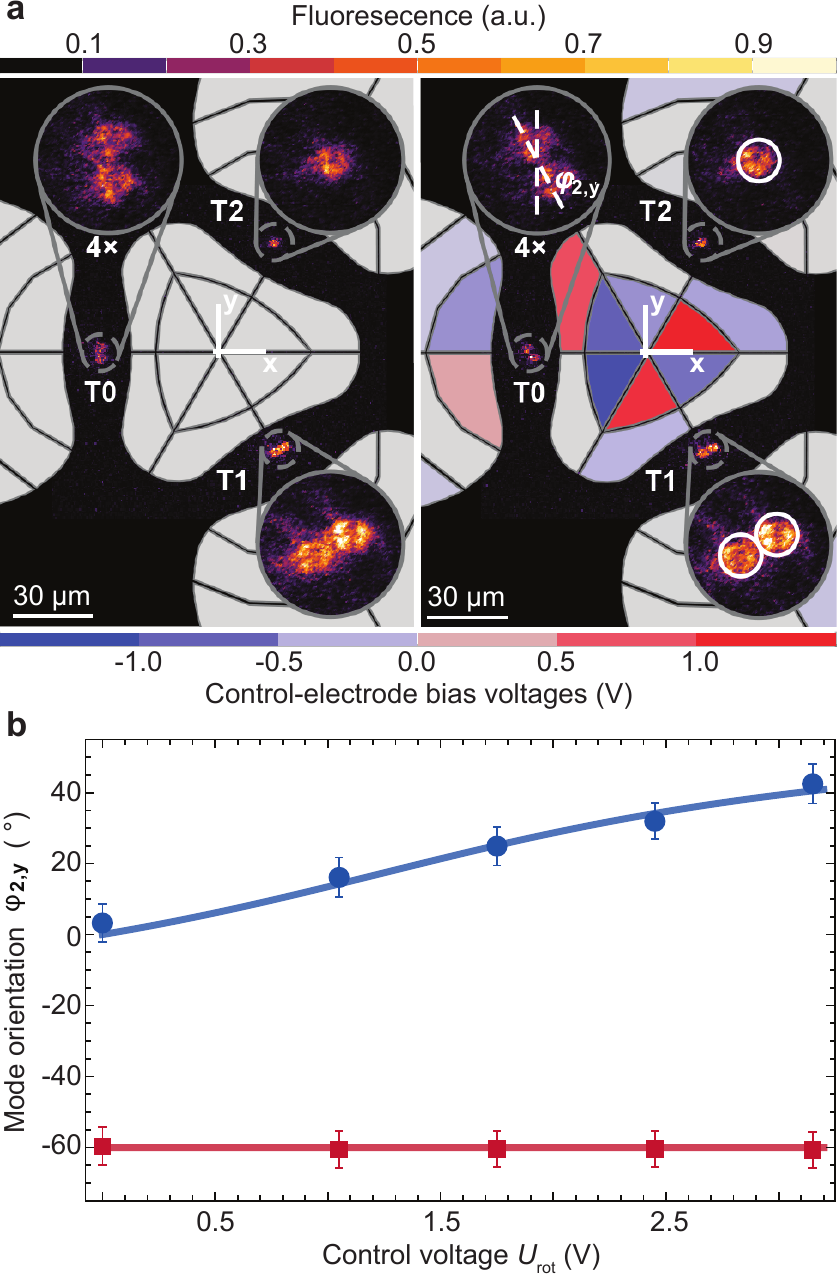}\\
   \caption{\textbf{Individual control of mode orientations.} %
     (\textbf{a}) EMCCD images of pairs of ions near $\mathbf{T0}$ and $\mathbf{T1}$, and a single ion near $
\mathbf{T2}$ (insets magnified by four).
Schematics of control electrodes are coloured according to their bias voltage $\mathbf{U}_{\rm{rot}}$.
Ion pairs align along $\mathbf{u}_2$, the lowest-frequency mode.
The left image captures ion positions for $U_{\rm{rot}}=0$\,V and, here, $\omega_2/(2\pi) = 1.9(2)$\,MHz near $\mathbf{T0}$.
The right image illustrates the rotation effect for $U_{\rm{rot}}=2.45$\,V:
Mode $\mathbf{u}_2$ near $\mathbf{T0}$ with $\omega_2/(2\pi) = 1.8(2)$\,MHz is rotated by $\varphi_{\textrm{2,y}} = 31(5)^\circ$, while ion positions near $\mathbf{T1}$ and $\mathbf{T2}$ remain unchanged; white circles indicate initial ion positions (for $U_{\rm{rot}}=0$).
    (\textbf{b}) Mode  $\mathbf{u}_2$ orientation in the $xy$ plane for $\mathbf{T0}$ (blue dots) and $\mathbf{T1}$ (red squares), described by $\varphi_{\textrm{2,y}}$, derived from a total of 14 images as a function of $U_{\rm{rot}}$; error bars denote our systematic uncertainty. The data is in good agreement with our predictions based on numerical simulation of $\kappa_{\rm{rot}}$ (solid lines).
   }\label{fig:rotationCCD}
\end{figure}

We rotate mode orientations near $\mathbf{T0}$ in the $xy$ plane with a control-potential $\hat\kappa_{\rm{rot}}$, while setting additional constraints to keep gradients and curvatures of the local trapping potential constant at $\mathbf{T1}$ and  $\mathbf{T2}$ (Methods).
We determine the rotation of mode orientations from EMCCD images of two ions near $\mathbf{T0}$ that align along $\mathbf{u}_2$ (axis of weakest confinement).
Simultaneously, we trap one or two ions near $\mathbf{T1}$ and $\mathbf{T2}$ to monitor residual changes in ion positions and mode orientations (and frequencies) due to unwanted local gradients and curvatures of $\hat\kappa_{\rm{rot}}$.
We take 14 images for five different $\hat{\kappa}_{\rm{rot}}$ values, while constantly Doppler cooling all ions and exciting fluorescence.
Figure~\ref{fig:rotationCCD}a shows two images for $U_{\rm{rot}} = 0$\,V (left) and $U_{\rm{rot}} = 2.45$\,V (right). Schematics of control electrodes are overlaid to the images and coloured to indicate their bias voltages $\mathbf{U}_{\rm{rot}}$.
%
Ion positions (in the $xy$ plane) are obtained with an uncertainty of $\pm 0.5\,\mu$m, yielding uncertainties for inferred angles $\varphi_{\textrm{2,y}}$ of $\pm 5^{\circ}$. Here, $\varphi_{\textrm{2,y}}$ denotes the angle between local mode $\mathbf{u}_2$ and  $y$. Figure~\ref{fig:rotationCCD}b shows measured $\varphi_{\textrm{2,y}}$ for ions near $\mathbf{T0}$ (blue dots) and $\mathbf{T1}$ (red squares) and compares them with our theoretical expectation (solid lines), further described in the Methods. We tune $\varphi_{\textrm{2,y}}$ between $0^{\circ}$ and $45^{\circ}$ near $\mathbf{T0}$, enabling us to set arbitrary mode orientations in the $xy$ plane, while ion positions (mode orientations) near $\mathbf{T1}$ and $\mathbf{T2}$ remain constant within $\pm 0.5\,\mu$m (better than $\pm 5\,^{\circ}$) in the $xy$ plane.

A complementary way of characterising mode orientations and frequencies, now with respect to $\mathbf{\Delta k}_{\textrm{x}}$ and/or $\mathbf{\Delta k}_{\textrm{y}}$ is to analyse the probability of finding $\ket{\downarrow}$ after applying $\ket{\downarrow} \leftrightarrow \ket{\uparrow}$ (carrier) or motional sideband couplings for variable duration. If all modes of a single ion are prepared in their motional ground state, the ratio of Rabi frequencies of carrier and sideband couplings is given by the Lamb-Dicke parameter~\cite{Leibfried2003a}, which is for $\mathbf{u}_1$ and $\mathbf{\Delta k}_{\textrm{x}}$:
\begin{equation}\label{Eq:LamDic}
\eta_{\rm{1,x}}=\mathbf{\Delta k}_{\textrm{x}}  \cdot \mathbf{u}_1 \sqrt{\frac{\hbar}{2\,m\,\omega_1}}=|\mathbf{\Delta k}
_{\textrm{x}}| \sqrt{\frac{\hbar}{2\,m\,\omega_1}}\cos(\varphi_{\textrm{1,x}})\textrm{,}
\end{equation}
where $\varphi_{\textrm{1,x}}$ is the angle between $\mathbf{u}_1$ and $\mathbf{\Delta k}_{\textrm{x}}$. The differences of carrier and sideband transition frequencies reveal the mode frequency, while ratios of carrier and sideband Rabi-frequencies determine Lamb-Dicke parameters and allow for finding the orientation of modes.

\begin{figure}[]
  \centering
    \includegraphics{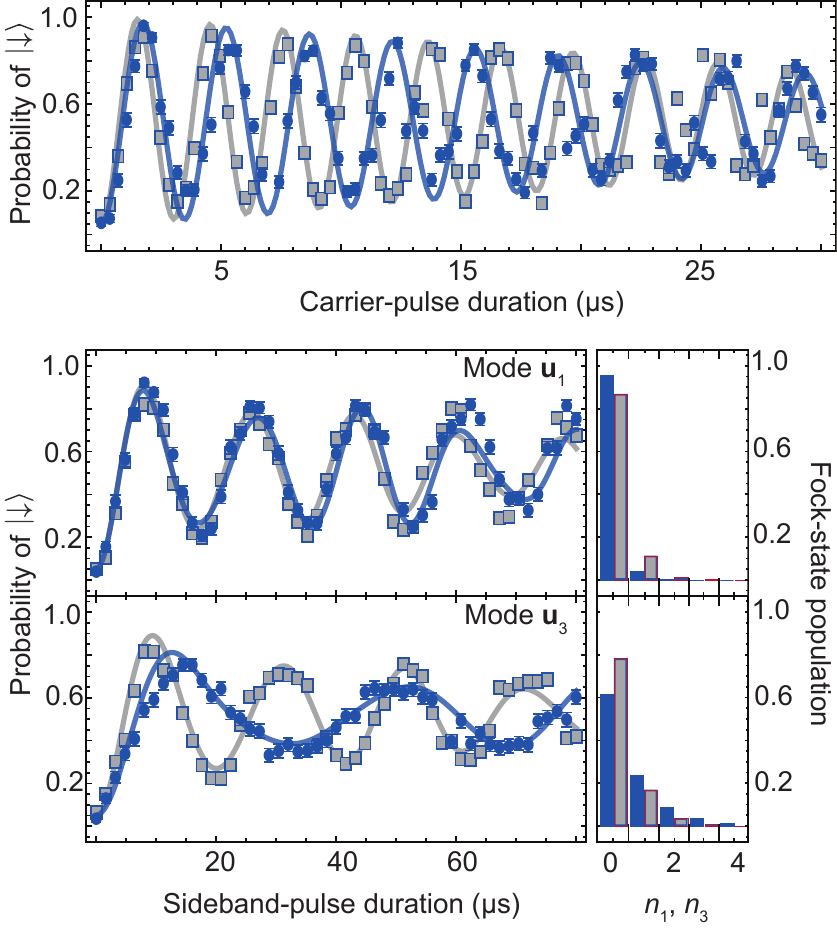}
   \caption{\label{fig:rotationRaman}
  \textbf{Measuring mode orientations via laser couplings.} Probability of $\ket{\downarrow}$ after applying carrier and sideband couplings via $\mathbf{\Delta k}_{\textrm{x}}$ with a single ion near $\mathbf{T0}$.
  Mode orientations are set with static $\hat\kappa_{\textrm{rot2}}$ (rotations in the $xz$ plane) for $U_{\rm{rot2}} = -1.62$\,V (blue dots) and $-2.43$\,V (grey squares). (Top) carrier transitions: $\ket{\uparrow, n_1, n_3} \leftrightarrow \ket{\downarrow, n_1, n_3}$, (middle) $\mathbf{u}_1$-sideband transitions: $\ket{\uparrow, n_1, n_3} \leftrightarrow \ket{\downarrow, n_1 +1, n_3}$, and (bottom) $\mathbf{u}_3$-sideband transitions: $\ket{\uparrow, n_1, n_3} \leftrightarrow \ket{\downarrow, n_1, n_3 +1}$.
  From combined model fits to all transitions (for each $\hat{\kappa}_{\rm{rot2}}$) we find angles $\varphi_{\textrm{1,x}} = 24.7(2)^\circ$ for $U_{\rm{rot2}} = -1.62$\,V (blue lines) and $36.1(2)^\circ$ for $U_{\rm{rot2}} = -2.43$\,V (grey lines) of mode $\mathbf{u}_1$ relative to $\mathbf{\Delta k}_{\textrm{x}}$.
  Histograms display derived Fock-state populations with thermal average occupation numbers between $\simeq 0.05$ and $\simeq 0.6$.
  Each datapoint is the average of 250 experiments and error bars (for some data smaller than symbols) denote the s.e.m.
   }
\end{figure}

We use a single ion near $\mathbf{T0}$ to determine the orientations and frequencies of two modes relative to $\mathbf{\Delta k}_{\textrm{x}}$.
We apply a control potential $\hat\kappa_{\textrm{rot2}}$, designed to rotate $\mathbf{u}_1$ and $\mathbf{u}_3$ in the $xz$ plane, and implement carrier and sideband couplings to both modes with $\mathbf{\Delta k}_{\textrm{x}}$ after resolved sideband cooling and initialising $\ket{\uparrow}$.
In Fig.~\ref{fig:rotationRaman}, the probability of $\ket{\downarrow}$ is shown for different pulse durations of carrier couplings (top) and sideband couplings on mode $\mathbf{u}_1$ (middle) and $\mathbf{u}_2$ (bottom).
Data points for $U_{\rm{rot2}} = -1.62$\,V are shown as blue rectangles and for $-2.43$\,V as grey rectangles.
We fit each data set to a theoretical model~\footnote{H. Kalis \emph{et al.} (unpublished)} (blue and grey lines) to extract the angles and distributions of Fock-state populations of each mode (shown as histograms):
We find $\varphi_{\textrm{1,x}}= 24.7(2)^\circ$ for $U_{\rm{rot2}} = -1.62$\,V and $\varphi_{\textrm{1,x}} = 36.1(2)^\circ$ for $U_{\rm{rot2}} = -2.43$\,V, while average occupation numbers range between $\simeq 0.05$ and $\simeq 0.6$.
Adding measurements along $\mathbf{\Delta k}_{\textrm{y}}$ and taking into account that the normal modes have to be mutually orthogonal would allow to fully reconstruct all mode orientations.
With resolved sideband cooling on all three modes, we can prepare a well defined state of all motional DoF.

\section*{Discussion}
We characterised two trap arrays that confine ions on the vertices of equilateral triangles with side lengths 80\,$\mu$m and 40\,$\mu$m.
We developed systematic approaches to individually tune and calibrate control potentials in the vicinity of each trap site of the 80 $\mu$m array, by applying bias potentials to 30 control electrodes.
With suitably designed control potentials, we demonstrated precise individual control of mode frequencies and orientations.
By utilising a multi-channel AWG, we also dynamically changed control potentials within single experimental sequences without adverse effects on spin or motional states.
Further, we devised a method to fully determine all mode orientations (and frequencies) based on the analysis of carrier and sideband couplings.
Measured heating rates are currently comparable to the expected inter-ion Coulomb coupling rate of $\Omega_\textrm{ex}/(2\pi) \simeq 1$\,kHz for $^{25}$Mg$^+$ ions in the 40 $\mu$m array at mode frequencies of $\simeq 2\pi\times 2$\,MHz~\cite{Brown2011}.
This coupling rate sets a fundamental time scale for effective spin-spin couplings~\cite{Wilson2014}.
To observe coherent spin-spin couplings, ambient heating needs to be reduced.
Decreases in heating rates of up to two orders of magnitude would leave $\Omega_\textrm{ex}$ considerably higher than competing decoherence rates and allow for coherent implementation of fairly complex spin-spin couplings. Such heating rate reductions have been achieved in other surface traps by treatments of the electrode structure~\cite{Allcock2011,Hite2012,daniilidis_surface_2014} and/or cryogenic cooling of the electrodes~\cite{deslauriers_scaling_2006,labaziewicz_suppression_2008,labaziewicz_temperature_2008}. The couplings in question have been observed in one dimension in a cryogenic system~\cite{Brown2011,Wilson2014}.

\begin{figure}[]
  \centering
  \includegraphics{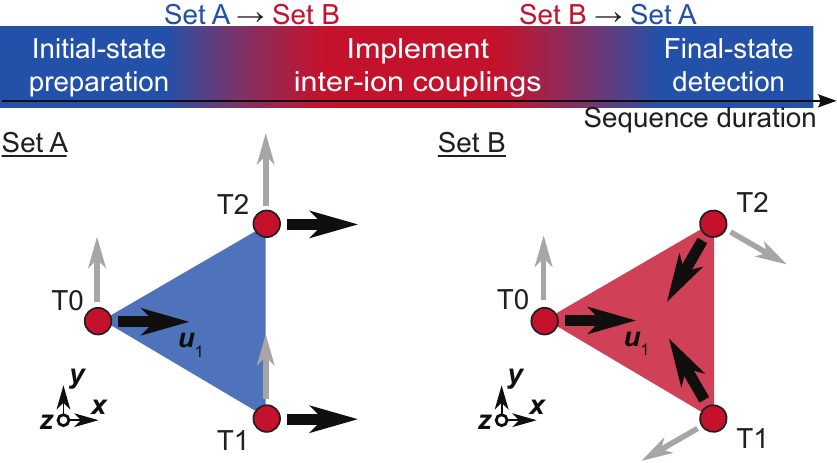}
  \caption{
  \textbf{Generic experimental sequence for tuneable inter-ion couplings.} %
  (Top) Time-line of an experiment starting with the initialisation of an fiducial quantum state of all ions in the array, followed by an adiabatic ramp of control potentials between sets A and B that reconfigure the normal mode structure from the setup shown on the bottom left to that shown on the bottom right.
  Then appropriate laser fields implements inter-ion couplings required for the AQS.
  After the simulation is complete, the potentials are a ramped back to set A, where the individual spin states can be detected.
  (Bottom) examples for configurations of motional DoF are illustrated by the arrows that show the orientation of $\mathbf{u}_1$ at the three trap sites (red dots).
  Set A may be applied when globally and/or locally preparing and detecting spin-motional states, while set B can establish specific inter-ion Coulomb couplings to mediate, e.g., effective spin-spin couplings for AQS, cp.~\cite{Schneider2012,Wilson2014}.
   }
  \label{fig:rotationRealTime}
\end{figure}

Currently we can compensate stray fields, set up normal mode frequencies and directions for all three ions and initialise them for a two-dimensional AQS, i.e., prepare a fiducial initial quantum state for ions at each trap site. A complete AQS may use the sequence presented in Fig.~\ref{fig:rotationRealTime}.
A dynamic ramp adiabatically transforms the system between two control sets, labelled A and B, that realise specific mode frequencies and orientations at each site.
Set A may serve to globally initialise spin-motional states of ions, potentially with more than one ion at each site, that could be the ground state of a simple initial Hamiltonian.
At all sites, mode frequencies and orientations need to be suitable (bottom left of Fig.~\ref{fig:rotationRealTime}) to enable global resolved sideband cooling, ideally preparing ground states for all motional modes.
A first ramp to set B combined with appropriate laser fields may be used to adiabatically or diabatically turn on complex spin-spin couplings.
Mode frequencies and orientations are tuned such that the Coulomb interactions between ions can mediate effective spin-spin couplings, e.g., all mode vectors $\mathbf{u}_1$ are rotated to point to the center of the triangle (bottom right of Fig.~\ref{fig:rotationRealTime}).
During the application of such interactions, the ground state of the uncoupled system can evolve into the highly entangled ground state of a complex coupled system.
Diabatic ramping to B will `quench' the original ground state and the coupled system will evolve into an excited state that is not an eigenstate.
After a final adiabatic or diabatic ramp back to set A, we can use global (or local) laser beams to read out the final spin states at each site.

In this way, our arrays may become an arbitrarily configurable and dynamically reprogrammable simulator for complex quantum dynamics.
It may enable, for example, the observation of photon assisted tunnelling, as required for experimental simulations of synthetic gauge fields~\cite{Bermudez2011,bermudez_photon-assisted-tunneling_2012} or other interesting properties of finite quantum systems, like thermalisation, when including the motional DoF~\cite{clos_time-resolved_2015}.
Concentrating on spin-spin interactions, the complex entangled ground states of spin frustration can be studied in the versatile testbed provided by arrays of individually trapped and controlled ions~\cite{Schmied2011,Schneider2012}.
Arrays with a larger number of trap sites could realise a level of complexity impossible to simulate on conventional computers~\cite{Shi2013,Nielsen2013}.

\section*{Methods}
\textbf{Architecture of the trap chip.}
 The $10 \times 10$\,mm$^2$ SiO$_2$ substrate of our trap chip is bonded onto a $33 \times 33$\,mm$^2$ ceramic pin grid array (CPGA); the electrodes of the trap arrays are wire-bonded with aluminum wires to the pins of the CPGA, with independent pins for the RF electrodes of the two arrays. The trap chip contains four aluminum-1/2\% copper metal layers, that are electrically connected by tungsten vertical interconnects thereby allowing `islanded' control electrodes in the top electrode layer (Fig.~\ref{fig:TrapSetup}). The buried electrical leads are isolated by intermediate SiO$_2$ layers, nominally 2\,$\mu$m thick, while the surface layer is spaced by $10\,\mu$m from the buried layers. All electrodes are mutually separated by nominally $1.2$ to $1.4$\,$\mu$m gaps and a 50 nm gold layer is evaporated on the top surfaces in a final fabrication step. The trap chip fabrication is substantially the same as that described in~\footnote{Section B of the Supplement to Ref.~\cite{Tabakov2015} herein}. Each control electrode is connected to ground by $860$\,pF capacitors located on the CPGA to minimise potential changes due to capacitive coupling to the RF electrodes.

\textbf{Minimising stray fields at each site.}
For compensation of local stray fields in the $xy$ plane, we vary the strength of individual control potentials $\hat\varepsilon_{\rm{x}}$ and $\hat\varepsilon_{\rm{y}}$ and find corresponding coefficient settings where we obtain a maximal Rabi rate of the detection transition and/or minimal Rabi rates of micromotion-sideband transitions probed with $\mathbf{\Delta k}_{\textrm{x}}$ and $\mathbf{\Delta k}_{\textrm{y}}$; resulting in residual stray-field amplitudes of $\leq 3\,$V\,m$^{-1}$. For compensation along  $z$, we vary the strength of individual $\hat\varepsilon_{\rm{z}}$ to minimise a change in ion position due to a modulation of $U_{\rm{RF}}$. The depth of field of our imaging optics aids to detect changes in $z$-position via blurring of images of single ions trapped at each site, within an uncertainty of about $\pm 5\,\mu$m. This corresponds to residual stray-field amplitudes of $\simeq 900$\,V\,m$^{-1}$ for typical trapping parameters.

\textbf{Mode frequency and heating rate measurements.} %
To measure mode frequencies, we Doppler-cool the ion and pump to $\ket{\downarrow}$. Then we apply a motional excitation pulse with fixed duration $t_\textrm{exc} = 100\,\mu$s to a single control electrode.
The pulse produces an electric field oscillating at a frequency $\omega_{\rm{exc}}$ that excites the motion, if $\omega_{\rm{exc}}$ is resonant with a mode frequency, and we can detect mode amplitudes of $> 100$\,nm along $\mathbf{k}_{\rm{D}}$ via the Doppler effect.
In the experiments, we vary $\omega_\textrm{exc}$ and obtain resonant excitations at $\omega_{j}$ with $j = \{1,2,3 \}$.
By repeating measurements, we record $\simeq 50$ consecutive frequency values for each mode frequency over the course of $\Delta t \simeq 1$\,h with a single ion near $\mathbf{T0}$. The results are consistent with a linear changes in frequencies, with corresponding slopes $\Delta \omega_{1}/{\Delta t} = - 1.50(5)\,$Hz\,s$^{-1}$, $\Delta \omega_{2}/{\Delta t} = - 1.07(2)\,$Hz\,s$^{-1}$, and $\Delta \omega_{3}/{\Delta t} = - 1.05(8)\,$Hz\,s$^{-1}$.

For the heating rate measurements, we add multiple resolved-sideband cooling pulses after Doppler cooling to our sequence and determine mode temperatures from the sideband-ratios for several different delay times~\cite{Turchette2000}.
In our experiments, we either use $\mathbf{\Delta k}_{\rm{x}}$ to iteratively address $\mathbf{u_1}$ and $\mathbf{u_3}$ or $\mathbf{\Delta k}_{\rm{y}}$ to address only $\mathbf{u_2}$.
For this, we prepare similar mode orientations as presented in Fig.~\ref{fig:rotationRaman}, find initial mode temperatures after cooling to $\bar{n}_{j} \lesssim 0.3$, and obtain corresponding heating rates.

\textbf{Potentials for individual control of motional DoF at each site.}
As a representative example for designing control potentials, we discuss $\hat\kappa_{\rm{rot}}$ that serves to rotate the normal modes in the $xy$ plane.
At position $\mathbf{T0}$ the constraints are:
\begin{equation}
[\partial_k \partial_l] \hat{\kappa}_{\textrm{rot}}(\mathbf{r})|_{\mathbf{r} = \mathbf{T0}} =
\left(
\begin{array}{ccc}
-1.60 & 1.75 & 0 \\
 1.75 & 0.84 & 0 \\
0 & 0 & 0.76 \\
\end{array}
\right)\times10^{7}\,\rm{m^{-2}}\textrm{, }
\label{eq:etarot}
\end{equation}
for $k$ and $l$ = $\{x, y, z\}$, while local gradients at all three trap sites and local curvatures at $\mathbf{T1}$ and $\mathbf{T2}$ are required to be zero.
We added diagonal elements in $[\partial_k \partial_l] \hat{\kappa}_{\textrm{rot}}(\mathbf{r})|_{\mathbf{r} = \mathbf{T0}}$ to reduce changes of the $\mathbf{u}_2$ frequency during variation of $\hat{\kappa}_{\rm{rot}}$ around our initial mode configurations. The mode configurations in the real array deviate from those derived from the $\phi_{\rm{ps}}$ due to additional curvatures near each trap site generated by stray potentials on our chip.
Ideally, we would design control potentials for mode rotations such that all frequencies stay fixed. This is only possible if we rotate around one of the mode vectors and know all initial mode vectors. Additionally, we keep mode vectors tilted away from $z$ to sufficiently Doppler cool all modes during state initialisation. Similarly, we design $\hat\kappa_{\rm{rot2}}$ to rotate modes in the $xz$ plane.

\textbf{Model for rotation of principal axes.} %
To model the rotation angle $\varphi_{2,y}$ of $\mathbf{u}_2$ near $\mathbf{T0}$  as a function of $\hat{\kappa}_{\rm{rot}}$, we consider the final trapping curvature at $\mathbf{T0}$ (analogously for neighbouring sites):
\begin{equation}
[\partial_k \partial_l] \phi_{\textrm{fin}}(U_{\rm{rot}})|_{\mathbf{r} = \mathbf{T}_0}\rm = 
[\partial_k \partial_l] \phi_{\textrm{ini}}|_{\mathbf{r} = \mathbf{T}_0}\rm{,}
+U_{\textrm{rot}}\,[\partial_k \partial_l] \kappa_{\textrm{rot}}|_{\mathbf{r} = \mathbf{T}_0}\rm{,}
\label{eq:curtot}
\end{equation}
where $\phi_{\textrm{ini}}(\mathbf{r})$ represents the initial potential, i.e., the sum of the pseudopotential, stray potential, and additional control potentials (used for stray field compensation).
The local curvatures (mode frequencies and vectors) of $\phi_{\textrm{ini}}(\mathbf{r})$ near $\mathbf{T0}$ are estimated from calibration experiments.
For simplicity, we reduce equation~(\ref{eq:curtot}) to two dimensions (in the $xy$ plane) and find corresponding eigenvectors and eigenvalues for $U_{\rm{rot}}$ between $0.0$\,V and $3.0$\,V.
We obtain angles $\varphi_{2,y}(U_{\rm{rot}})$ of the eigenvector $\mathbf{u}_2$ and we show resulting values as an interpolated solid line in Fig.~\ref{fig:rotationCCD}b.

\section*{References}

\section*{Acknowledgements}
This work was supported by DFG (SCHA 972/6-1). Sandia National Laboratories is a multiprogram laboratory managed and operated by Sandia Corporation, a wholly owned subsidiary of Lockheed Martin Corporation, for the U.S. Department of EnergyÕs National Nuclear Security Administration under Contract No. DE-AC04-94AL85000. %
We thank D. M\"ohring for fruitful discussions and J. Denter
for technical assistance. Further, we are grateful for helpful comments on
the manuscript given by S.~Todaro, K.~McCormick, and Y.~Minet. %

\section*{Author contributions}
M.M., H.K., and U.W. participated in the design of the experiment and built the experimental apparatus. %
M.M, H.K., M.W., F.H, and U.W. collected data and analysed results. %
M.M., U.W., D.L., and T.S. wrote the manuscript. %
R.S. and D.L. participated in the design of the trap arrays and the experiment. %
M.B. and P.M. participated in the design and fabricated the trap chips. %
T.S. participated in the design and analysis of the experiment. %
All authors discussed the results and the text of the manuscript.%

\section*{Additional information}
Competing financial interests: The authors declare no competing financial interests. %

\end{document}